\documentclass[%
reprint,
 amsmath,amssymb,
 aps,
]{revtex4-2}

\usepackage[utf8]{inputenc}
\usepackage{graphicx}
\usepackage{bm}

\begin{document}

\preprint{APS/123-QED}

\title{Unconventional resistive switching in dense Ag-based nanowire networks with brain-inspired perspectives}


\author{Juan I. Diaz Schneider, Eduardo D. Mart{\'i}nez, Pablo E. Levy}
\affiliation{%
Consejo Nacional de Investigaciones Cient{\'i}ficas y T{\'e}cnicas (CONICET), Argentina.\\
Instituto de Nanociencia y Nanotecnolog{\'i}a (CNEA - CONICET), Nodo Bariloche.\\
Gerencia F{\'i}sica, Centro At{\'o}mico Bariloche, Comisi{\'o}n Nacional de Energ{\'i}a At{\'o}mica (CNEA), Av. Bustillo 9500, (8400) S. C. de Bariloche, R{\'i}o Negro, Argentina. \\
}%

\author{Cynthia P. Quinteros}
 \email{cquinteros@unsam.edu.ar}
\affiliation{%
Instituto de Ciencias F{\'i}sicas (UNSAM-CONICET), Mart{\'i}n de Irigoyen 3100, San Mart{\'i}n (1650), Argentina.
}%

\date{\today}

\begin{abstract}
\noindent  
We report an unconventional resistive switching effect on high-density self-assembled Ag-nanowire networks tailored by a fuse-like operation.
Analyzing the electrical signatures, before and after such a fusing, we propose a mechanism to rationalize the observed phenomenology.
The explanation allows to reconcile the results obtained in similar systems early adopted as transparent electrodes and the more recent attempts to use this type of substrate for \textit{in-materia} computational operations. 
In addition to the usual analog-nature of the available resistance states and the ability to tune internal weights, here we show that sparsity and non-linear behavior are also attributes of these networks. 
Thus, the formerly exhibited nanowires' abilities to code synaptic behavior are complemented by neuronal features upon properly tuning the network density and the applied electrical protocol. 
\end{abstract}

\maketitle

\noindent Self-assembled networks (SANs) formed by metallic nanowires (NW) are intensively studied as a bio-inspired playground for a neuromorphic alternative to the traditional digital computation approach \cite{markovic_physics_2020,kuncic_neuromorphic_2021}. In this architecture, some features of their biological counterparts are achieved: electrical conductivity \cite{sannicolo_direct_2016}, tunability of internal resistance states \cite{diaz_schneider_resistive_2022}, emergence of collective properties \cite{di_francesco_spatiotemporal_2021}, and a high degree of interconnectivity \cite{chialvo_emergent_2010}. Looking closer at the constituents, metallic nanowires may be compared to the biological axons while their coating forms capacitor-like structures at the crossing points between them. The ability to tune the resistance at the cross-point of two coated NWs, in a memristive-like way, is comparable to adapting the synaptic weight between two neighboring neurons. Moreover, the connectivity of the nanowires -conditioned by synthesis and electrical measurements \cite{diaz_schneider_two-junction_nodate}- determines the intrinsic self-organization of the network whose conductivity and, consequently, overall resistance is governed by the number of available paths connecting the external electrodes. 

As we have previously reported for Ag-based SANs, different strategies could be used to trigger their response towards desired new functionalities: NWs density (during synthesis) \cite{diaz_schneider_two-junction_nodate}, environmental conditions \cite{diaz_schneider_resistive_2022}, and electrical protocol. In our first contribution \cite{diaz_schneider_resistive_2022}, the role of humidity was emphasized and preliminarily explored. Lately, we have systematically analyzed the electrical response of Ag-based SANs covering the full percolation range and offering a generalized model to reproduce all the observed behaviors \cite{diaz_schneider_two-junction_nodate}. In this work, we focus on high-density samples and the underlying mechanism to explain their pristine low-resistance state that after a fuse-like operation becomes more resistive and switchable. 
The detailed study of these SANs and the proposed scenario allow to interpret the experimental data and to reconcile the transport properties of the NWs-based networks formerly used as transparent electrodes and those more recently exploited as neuromorphic hardware alternatives. Moreover, here we extend the characterization to argue that many of the primitives requested to implement brain-inspired computing \cite{kendall_building_2020-1} are fulfilled, and to postulate these high-density AgNWs-based SANs as interesting candidates for more than just synaptic functionalities.

\vspace{0.1cm}

Silver nanowires (Ag-NWs) coated with polyvinylpyrrolidone (PVP) were fabricated following the standard polyol synthesis \cite{jiu_facile_2014}. Obtained NWs are 10$^{-4}$ m long and 10$^{-7}$ m in diameter, as observed using Scanning Electron Microscopy \cite{diaz_schneider_resistive_2022}. SANs were fabricated following spin coating \cite{khanarian_optical_2013} and multi-deposition steps \cite{zhu_convertible_2019-1}. Dense SANs, i.e. those fabricated well beyond the electrical percolation limit \cite{diaz_schneider_two-junction_nodate}, were produced. The samples consist of SANs spontaneously formed after spreading the Ag-NWs, the latter coated by the remaining polymer employed during synthesis. This implies that the Ag-NWs are not directly exposed to the environment but protected by an ultrathin PVP-layer \cite{diaz_schneider_resistive_2022}. Moreover, the presence of PVP at the NWs cross-points is believed to be key to the availability of multiple internal resistance states (plasticity  \cite{kendall_building_2020-1}) within these SANs. 

Glass substrates with previously deposited Ag pads (two parallel electrodes 1 mm wide being 1 mm apart) were used. External wires were attached to the electrodes using silver paste. Electrical measurements were performed using a Keithley SMU 2450 unit. A 2-wire method was used by applying 'V' and measuring 'I'. 
The ambient relative humidity was controlled to be around 33 $\%$ \cite{diaz_schneider_resistive_2022}. 

\vspace{0.3cm}

\begin{figure}[ht!]
        \centering
        \includegraphics[width=\columnwidth]{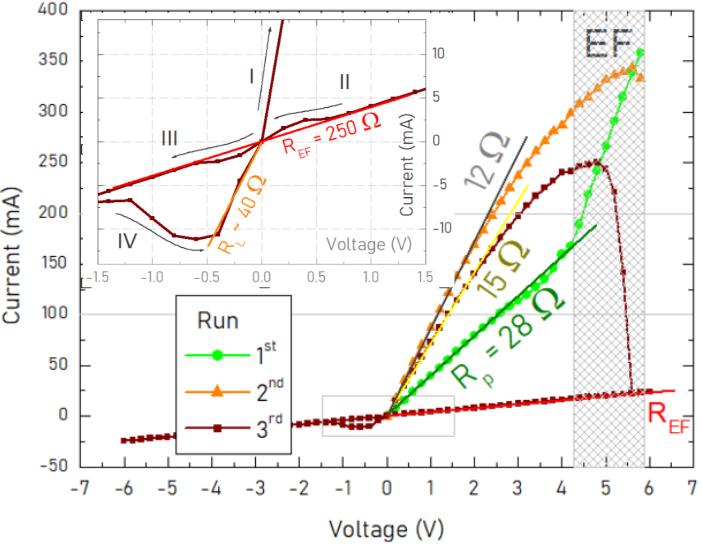}
        \caption{\label{fig:F1} \textbf{
        Initial stages of the tailoring process towards unconventional RS.} The pristine state (R$_{\mathrm{p}} \sim$ 30 $\Omega$) remains unaltered below 100 mA. Further increasing and repeating the voltage sweep produces a smooth reduction and ulterior marked increase of R (Runs 1$^{\mathrm{st}}$ and 2$^{\mathrm{nd}}$). 
        During the 3$^{\mathrm{rd}}$ Run a marked increase of R is observed close to +5 V (EF process) which determines a linear behavior when V is decreased. 
        This newly achieved linear regime (R$_{\mathrm{EF}}$ $\sim$ 250 $\Omega$) displays hints of different behavior in the low voltage range -2 V $<$ V $<$ +2 V (see the inset). \underline{Inset}: Zoom-in on the 3$^{\mathrm{rd}}$ Run. The R $\sim$ 15 $\Omega$ observed before EF (path I) become R$_{\mathrm{EF}}$ $\sim$ 250 $\Omega$ (path II). In the proximity of $\pm$ 0.5 V, while approaching 0 V (path II) and departing from 0 V (path III) a tiny swing of the current from linearity is observed. Finally, the way back from -6 V 
        (path IV) depicts a resistance change (R$_{\mathrm{L}} \sim$ 40 $\Omega$) which will be representative of the ulterior switching ability (see Figure \ref{fig:F2}).}     
\end{figure}

SANs with a dense population of Ag-NWs exhibit an ohmic-like (i.e. linear I-V response) low-resistance (R$_{\mathrm{p}} = \frac{\Delta \mathrm{V}}{\Delta \mathrm{I}}$ $\sim$ 30 $\Omega$) pristine state. Although the PVP coating of NWs should impede conduction at NW-NW cross-points, no threshold voltage was detected (see Run 1 in Figure \ref{fig:F1}). R$_{\mathrm{p}}$ remains almost constant below 100 mA upon increasing voltage. 

The low resistance of the pristine state and the absence of a threshold to enable conduction are difficult to rationalize considering the material constituents. The two characteristics are probably related. Despite the NWs being coated by a polymeric layer, the I-V response is linear, behaving as if the metal-insulator-metal structures at the NWs junctions would not exist. Moreover, the pristine resistance corresponds to the value expected for a unique NW (eq. \ref{eq:RNW} considers the mean values reported in \cite{diaz_schneider_resistive_2022}).

\begin{equation}
    \mathrm{R_{NW}} = \frac{\rho_{Ag} \cdot L}{area} \sim \frac{10^{-8} ~\Omega ~m \cdot 70 ~\mu m}{\pi \cdot \frac{(170 ~nm)^2}{4}} \sim 30 ~\Omega
    \label{eq:RNW}
\end{equation}

\noindent 
Here we argue that several junctions could be short-circuited by defective polymeric layers containing pin-holes and remnant Ag ions which, in turn, enable the conduction in an otherwise insulating material. Complementarily, the huge amount of percolative paths (dense SANs could comprise up to 3000 $\frac{\mathrm{NWs}}{\mathrm{mm}^2}$) determines a combination of series and parallel elements each of them having contributions from the metallic NWs (R$_{\mathrm{NW}}$) and the junctions (R$_{\mathrm{J}}$). A network consisting of many sites $i$ with R$_i$ = R$_{\mathrm{NW},i}$ + R$_{\mathrm{J},i}$ each, in arbitrary series / parallel connection, leads to an effective R$_{\mathrm{eff}}$ = $\gamma \cdot$ R$_i$, where $\gamma$ is a constant of the order of 1. Its precise value is related to the network's topology \cite{venezian_resistance_1994,cserti_application_2000}. 

The experimentally measured value of our pristine samples is R$_{\mathrm{p}}$ $\sim$ 10 - 30 $\Omega$, in quantitative agreement with the estimation R$_{\mathrm{NW}}$ of a single NW (30 $\pm$ 16 $\Omega$, considering the length's and diameter's mean values with their standard deviations \cite{diaz_schneider_resistive_2022}). Thus, as R$_{\mathrm{eff}}$ = $\gamma \cdot$ R$_i$ = $\gamma \cdot$ (R$_{\mathrm{NW},i}$ + R$_{\mathrm{J},i}$) $\sim$ R$_{\mathrm{NW}}$, we conclude that the contribution of each junction R$_{\mathrm{J},i}$ is small or negligible at the pristine stage. In this context, conducting junctions (having a low R-value) shortcut the contribution from higher R memristive junctions, and the main contribution to R$_{\mathrm{eff}}$ comes from Ag-NWs. Thus, in the pristine state, the collective response (i.e. macroscopic  R$_{\mathrm{p}}$ = R$_{\mathrm{eff}}$) is qualitative and quantitatively similar to the correspondent magnitude of the individual comprising element (i.e. microscopic R$_{\mathrm{NW},i}$). Our results are in agreement with findings by Bellew \textit{et al.} \cite{bellew_resistance_2015}, regarding the fact that the R$_{\mathrm{J}}$ contribution can be comparable to or less than that of R$_{\mathrm{NW}}$.  

In the range 100 mA - 200 mA some samples display a slight decrease of R (1$^{\mathrm{st}}$ Run in Figure \ref{fig:F1}), which could be related to optimizing the junction transport capability \cite{sannicolo_electrical_2018}. During the 2$^{\mathrm{nd}}$ Run, for currents exceeding 200 mA, the response smoothly departs from linearity, and R increases, an effect probably related to the temperature increase of the metallic NWs. 
During the 3$^{\mathrm{rd}}$ Run, a marked R increase becomes evident, hereafter termed electro-fusing (EF) \cite{diaz_schneider_two-junction_nodate}. The sequence is described as an optimization-degradation-breakdown process of the conducting network \cite{sannicolo_electrical_2018}.


%
Due to the geometry of the as-obtained network, both resistive and memristive junctions that were initially bypassed by the fusing current will afterward contribute to determining the R$_{\mathrm{EF}}$ value. After NW fusing, a reconfiguration of the current distribution occurs. Thus, although changes at the memristive junctions are prone to occur \cite{milano_brain-inspired_2020}, in this scenario the increase from R$_{\mathrm{p}}$ to R$_{\mathrm{EF}}$ appears to be mostly related to a 
topological change of the network.

Plenty of independent percolative paths are available in the pristine sample (this is related to the areal density of NWs and the lateral dimensions of the electrodes), and a constant R-value is obtained at low stimulus. 
Eventually, when high enough electrical power is applied (to the partially degraded network), a self-selective path disruption is produced, which determines an abrupt R increase \cite{sannicolo_electrical_2018} reminiscent of a fuse device. The observed EF process is non-polar and irreversible \cite{diaz_schneider_resistive_2022}. Irreversibility may arise on our experimental conditions (electrodes' separation, SANs density, or applied electrical protocol). Other authors have shown a more controlled disruption for individual NWs that can be rewired back upon proper electrical stimulation\cite{milano_electrochemical_2024-1}. It is also extremely dependent on the sweep velocity as previously studied in detail by Kholid \textit{et al.} \cite{kholid_multiple_2015}. EF is a unique process that configures the NW network. Note that the increase of R produced during this anti-percolative process \cite{sannicolo_direct_2016} is related to the lack of preferential pathways, which at first sight could be assimilated to a decrease in the effective SAN's density. However, this decrease in the number of available conducting paths is self-selected during the EF process, meaning it is not homogeneous nor uniform as a coverage or density change would be (i.e. a tailored sparsity is attained \cite{kendall_building_2020-1}). 

\begin{figure}[ht!]
        \centering
        \includegraphics[width=\columnwidth]{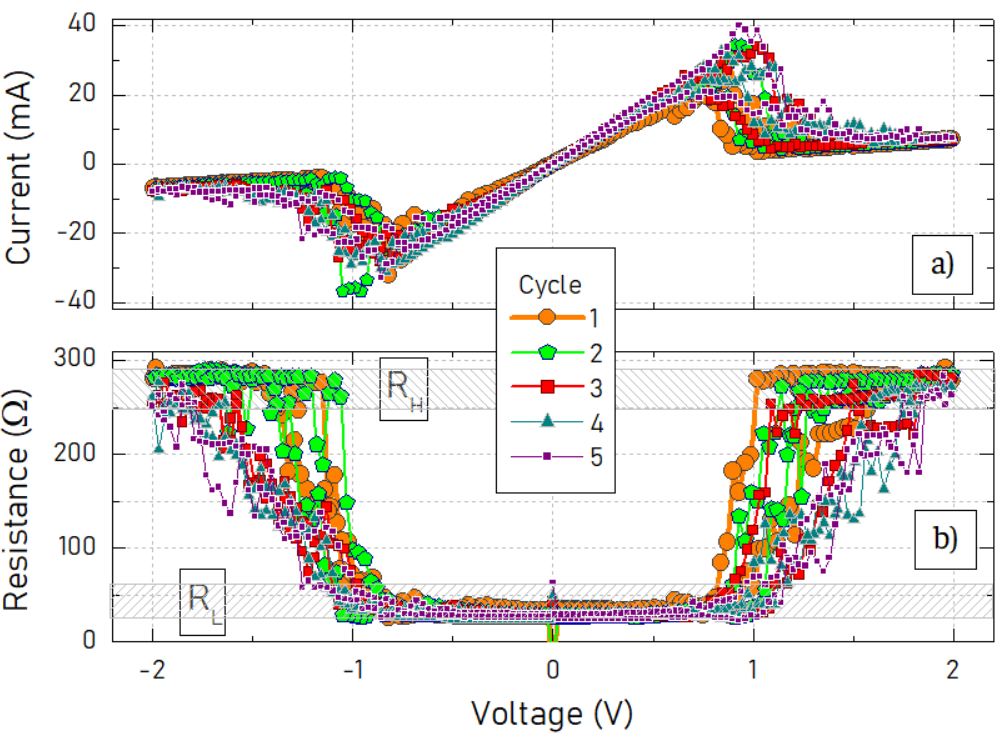}
        \caption{\label{fig:F2} \textbf{Unconventional RS in dense SANs after EF.} a) Current and b) resistance response upon voltage cycling in the $\pm$2 V range (same sample as in Figure \ref{fig:F1}). While ramping up the voltage, R$_{\mathrm{L}}$ $\sim$ 40 $\Omega$ transitions to R$_{\mathrm{H}}$ $\sim$ 250 $\Omega$ in the proximity of V$_{\mathrm{S}}$ $\sim$ $\pm$1 V. Upon reducing the voltage amplitude, R$_{\mathrm{L}}$ $\sim$ 40 $\Omega$ is recovered while crossing approximately the same V$_{\mathrm{S}}$ condition.}
\end{figure}

Upon further cycling with reduced-power voltage sweeps in the $\pm$2 V range (0 V to $\pm$2 V back to 0 V), a constant R$_{\mathrm{EF}}$ response (i.e. linear I-V characteristics) was observed until, unexpectedly, another constant resistance R$_{\mathrm{L}}$, lower than the former, was measured. The inset of Figure \ref{fig:F1} depicts this situation, displaying R$_{\mathrm{L}} \sim$ 40 $\Omega <$ R$_{\mathrm{EF}}$ ($\sim$ R$_{\mathrm{H}}$) = 250 $\Omega$. Cycling the sample confirms that R$_{\mathrm{L}} \sim$ 40 $\Omega$ is repeatedly observed at low stimulus for both polarities, as shown in Figure \ref{fig:F2}. Upon increasing the applied voltage, close to +1 V ($\pm$ 0.2 V), a marked current decrease is observed, and the system recovers the R$_{\mathrm{EF}}$ value. This RS effect, which we call unconventional, is non-polar, exhibits a resistance ratio of 6, and is produced repeatedly at a switching voltage V$_{\mathrm{S}}$ $\sim$ $\pm$ 1 V. 

\begin{figure}[ht!]
        \centering
        \includegraphics[width=\columnwidth]{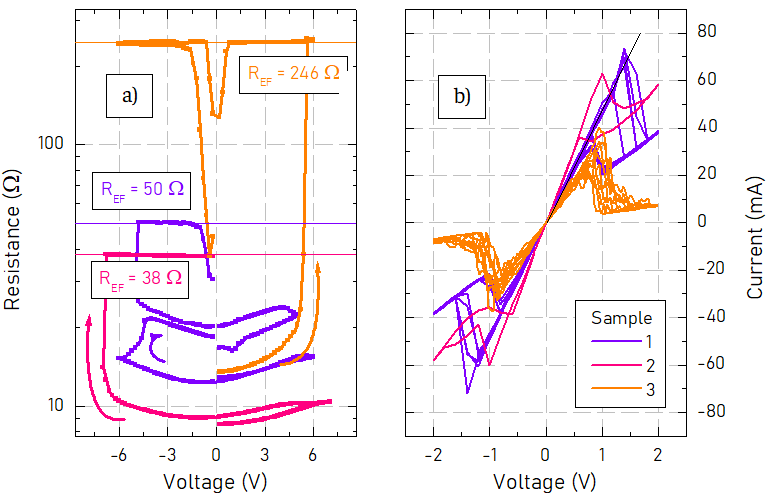}
        \caption{\label{fig:F3} \textbf{EF and unconventional RS in multiple samples.} a) Resistance as a function of voltage during EF for three dense SANs. b) Ulterior switching (after EF) corresponding to the same SANs illustrated in a).}
\end{figure}

Further insight is obtained by studying other dense samples. Figure \ref{fig:F3}a shows the precise resistance change upon EF for three samples. The selection of curves indicates the electro-fusing operation's non-polar nature, which could be triggered upon positive or negative voltages. Also, a partial EF process is identified among them. Interestingly, regardless of the details of the EF process, the three samples depict the ulterior switching at low voltage described in Figure \ref{fig:F2} (see Figure \ref{fig:F3}b). Its main feature is that 
a remnant low-resistance state (R$_{\mathrm{L}}$) becomes higher (R$_{\mathrm{H}}$) in the proximity of V$_{\mathrm{S}}$ (in this case $\sim \pm$1 V). This is observed regardless of the voltage polarity. Moreover, upon reducing the stimulus from $\pm$ 2 V to 0 V, the low-resistance state is recovered near the same V$_{\mathrm{S}}$ value. As observed in Figure \ref{fig:F3}b, the specific values for R$_{\mathrm{L}}$ and  R$_{\mathrm{H}}$ and the switching condition V$_{\mathrm{S}}$ may vary from sample to sample and upon cycling but they maintain within the range R$_{\mathrm{p}} <$ R$_{\mathrm{L}} <$ R$_{\mathrm{H}} \leq~$ R$_{\mathrm{EF}}$ and  V$_{\mathrm{S}} \sim \pm 1$ V.  

After EF, the low-resistance defective NW-PVP-NW junctions subjected to high-power dissipation have been removed, and other more resistive ones start to carry part of the current flow. 
Therefore, memristive junctions (comprising polymeric layers of insulating media) govern the electrical behavior. As demonstrated by other authors \cite{yang_electrochemical_2014,wang_surface_2019-1,milano_brain-inspired_2020,daniels_reservoir_2022}, conducting filaments (CF) form within the insulating nature. This forming voltage could be extremely low due to the availability of Ag ions and the intrinsic built-in electric field due to the NWs' curvature. 


We have previously shown \cite{diaz_schneider_two-junction_nodate} that a picture comprising low-resistance defective junctions getting wiped out upon EF and the insulating junctions, prone to filament formation, becoming dominant after EF is a successful model to explain the results. In this scenario, after EF, the samples describe a linear I-V regime until V$_{\mathrm{S}}$ when an RS to another linear regime is observed. The switching mechanism is a competition between the electric field, to hold the conducting filaments in place, and the power dissipation, which tends to dissolve them. Eventually, the heat becomes intense enough to disconnect some conducting filaments producing a resistance increase. 
Thereafter, the electric field at the junction promotes the restitution of the filament. Being a non-deterministic process, there is an inherent uncertainty regarding whether switching occurs at voltages higher than or lower than V$_{\mathrm{S}}$. 
In either case, there exists a voltage condition, close to V$_{\mathrm{S}}$, where switching is more likely, and this could be used to implement certain neuronal functionalities. 

\begin{figure}[ht!]
        \centering
        \includegraphics[width=\columnwidth]{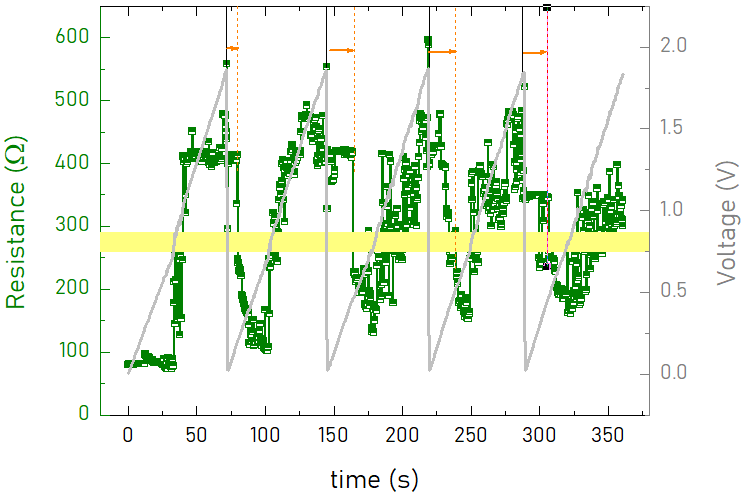}
        \caption{\label{fig:F5} \textbf{Threshold switching after EF.} Resistance and voltage as a function of time (left and right axis, respectively) of a dense SAN after EF. While ramping up the applied voltage and upon surpassing the threshold voltage (highlighted), R$_{\mathrm{L}}$ switches to R$_{\mathrm{H}}$. When the voltage is turned off, R$_{\mathrm{H}}$ persists during a relaxation process.}
\end{figure}

After the EF, the presence of a threshold which we have named after V$_{\mathrm{S}}$ implies a sharp transition from R$_{\mathrm{L}}$ to R$_{\mathrm{H}}$ that seems to vanish after a certain time lapse, or at least does not seem to hold when the applied voltage is turned off. To illustrate to what extent this is comparable to the soma behavior, Figure \ref{fig:F5} displays the switching we have been describing but conveniently plotted as a function of time. The externally applied stimulus consists of a train of sweeps of increasing voltage from 0 to +2 V. After reaching the maximum value, the voltage is turned off, and the positive sweep restarts. Each time the stimulus crosses the threshold value, resistance increases from tens (R$_{\mathrm{L}}$) to hundreds (R$_{\mathrm{H}}$) of $\Omega$. Further increasing the external voltage, does not significantly changes the resistance. 
Upon turning off the applied voltage, the achieved state is retained for a lapse before returning to its previous low resistance value.

%

Formatted in this way, it is possible to relate the characteristics of the unconventional switching identified in these dense AgNW SANs with the one attributed to the soma in biological neural tissues \cite{dayan_theoretical_2005}: existence of a threshold upon which the neuron reacts temporarily changing its resistance state. The procedure in biological entities is as follows: the soma holds inactive until a certain threshold of the cumulative signals collected in all the incoming connections is reached. Upon surpassing the threshold, the resistance changes. This is known as the fire function. The resistance change is strongly non-linear, in clear contrast with the accumulation change desirable for synaptic implementations. After the soma has reacted, it progressively relaxes and remains idle to any other stimulus. Eventually, the electrical signal produced by the soma decays recovering the original state before the switching. Similarities between the biological behavior and the observed experimental response reported here allows to recognize the potential of these systems to implement much more than just synaptic abilities.

Overall, these high-density AgNWs-based SANs reveal as promising building blocks for brain-inspired computing since their features comprise multiple of the primitives considered to be key for neuromorphic applications \cite{kendall_building_2020-1}. Starting from the self-assembled nature of the system and combined with the high-density of the AgNW population, a high degree of connectivity is intrinsically assured. Noteworthy, sparsity can be achieved using the EF described here. Complementarily, plasticity and analog behavior (the ability to tune synaptic weights, and adopt multiple resistance values, respectively) have been thoroughly analyzed in previous studies \cite{diaz_schneider_two-junction_nodate,milano_brain-inspired_2020,zhu_information_2021}. Finally, the observation of the threshold switching, here referred to as unconventional, demonstrates a non-linear electrical response, postulating the system to implement the neuron firing function. Although self-assemblies of NWs are usually thought of as \textit{in-materia} implementations of the internal layers of artificial neural networks (with the neuron function implemented via CMOS designs), the unconventional switching reported here offers a mechanism to implement the non-linear behavior expected for a synthetic neuron.

In summary, the electric transport response of dense self-assembled networks formed by coated silver nanowires was investigated. Their over-percolated nature determines a pristine low resistance. 
A controlled electro-fusing process selectively disrupts the conducting network, increasing its resistance but retaining a percolative character. Electrofusing dense samples reduces the number of available low-resistance paths, bringing the system closer to the percolation limit in an externally controlled way. Electrically induced breakdown of nanowires -a major problem when networks are used as transparent electrodes- is shown to be a controlled way of reducing the connectivity and obtaining unique samples exhibiting a resistive switching comparable to some extent to the neuron's firing function. Identifying this set of features allows us to widen the range of plausible applications of experimental realizations based on AgNW SANs. 

\section*{Data Availability Statement}
The data that support the findings of this study are available from the corresponding author upon reasonable request.

\begin{acknowledgments}
\noindent We are grateful to J. Lipovetzky (CAB-CNEA) for technical assistance and Prof. E. Miranda and J. Nino for valuable discussions. This work was supported by ANPCyT-FONCyT PICT-2021-I-A-00876 ('NNA') and CONICET PIP 11220220100508CO ('RdM'). EDM acknowledges financial support from CONICET PIBAA 2022-2023 28720210100473CO. CPQ acknowledges financial support from CONICET PIBAA 2022-2023 28720210100975CO and PICT INVI 2022 01097. JIDS acknowledges a fellowship from CONICET.
\end{acknowledgments}

\bibliographystyle{apsrev4-1}
\bibliography{Letter}

\begin{thebibliography}{26}%
\makeatletter
\providecommand \@ifxundefined [1]{%
 \@ifx{#1\undefined}
}%
\providecommand \@ifnum [1]{%
 \ifnum #1\expandafter \@firstoftwo
 \else \expandafter \@secondoftwo
 \fi
}%
\providecommand \@ifx [1]{%
 \ifx #1\expandafter \@firstoftwo
 \else \expandafter \@secondoftwo
 \fi
}%
\providecommand \natexlab [1]{#1}%
\providecommand \enquote  [1]{``#1''}%
\providecommand \bibnamefont  [1]{#1}%
\providecommand \bibfnamefont [1]{#1}%
\providecommand \citenamefont [1]{#1}%
\providecommand \href@noop [0]{\@secondoftwo}%
\providecommand \href [0]{\begingroup \@sanitize@url \@href}%
\providecommand \@href[1]{\@@startlink{#1}\@@href}%
\providecommand \@@href[1]{\endgroup#1\@@endlink}%
\providecommand \@sanitize@url [0]{\catcode `\\12\catcode `\$12\catcode `\&12\catcode `\#12\catcode `\^12\catcode `\_12\catcode `\%12\relax}%
\providecommand \@@startlink[1]{}%
\providecommand \@@endlink[0]{}%
\providecommand \url  [0]{\begingroup\@sanitize@url \@url }%
\providecommand \@url [1]{\endgroup\@href {#1}{\urlprefix }}%
\providecommand \urlprefix  [0]{URL }%
\providecommand \Eprint [0]{\href }%
\providecommand \doibase [0]{http://dx.doi.org/}%
\providecommand \selectlanguage [0]{\@gobble}%
\providecommand \bibinfo  [0]{\@secondoftwo}%
\providecommand \bibfield  [0]{\@secondoftwo}%
\providecommand \translation [1]{[#1]}%
\providecommand \BibitemOpen [0]{}%
\providecommand \bibitemStop [0]{}%
\providecommand \bibitemNoStop [0]{.\EOS\space}%
\providecommand \EOS [0]{\spacefactor3000\relax}%
\providecommand \BibitemShut  [1]{\csname bibitem#1\endcsname}%
\let\auto@bib@innerbib\@empty
\bibitem [{\citenamefont {Markovic}\ \emph {et~al.}(2020)\citenamefont {Markovic}, \citenamefont {Mizrahi}, \citenamefont {Querlioz},\ and\ \citenamefont {Grollier}}]{markovic_physics_2020}%
  \BibitemOpen
  \bibfield  {author} {\bibinfo {author} {\bibfnamefont {D.}~\bibnamefont {Markovic}}, \bibinfo {author} {\bibfnamefont {A.}~\bibnamefont {Mizrahi}}, \bibinfo {author} {\bibfnamefont {D.}~\bibnamefont {Querlioz}}, \ and\ \bibinfo {author} {\bibfnamefont {J.}~\bibnamefont {Grollier}},\ }\href {\doibase 10.1038/s42254-020-0208-2} {\bibfield  {journal} {\bibinfo  {journal} {Nature Reviews Physics}\ }\textbf {\bibinfo {volume} {2}},\ \bibinfo {pages} {499} (\bibinfo {year} {2020})}\BibitemShut {NoStop}%
\bibitem [{\citenamefont {Kuncic}\ and\ \citenamefont {Nakayama}(2021)}]{kuncic_neuromorphic_2021}%
  \BibitemOpen
  \bibfield  {author} {\bibinfo {author} {\bibfnamefont {Z.}~\bibnamefont {Kuncic}}\ and\ \bibinfo {author} {\bibfnamefont {T.}~\bibnamefont {Nakayama}},\ }\href {\doibase 10.1080/23746149.2021.1894234} {\bibfield  {journal} {\bibinfo  {journal} {Advances in Physics: X}\ }\textbf {\bibinfo {volume} {6}},\ \bibinfo {pages} {1894234} (\bibinfo {year} {2021})}\BibitemShut {NoStop}%
\bibitem [{\citenamefont {Sannicolo}\ \emph {et~al.}(2016)\citenamefont {Sannicolo}, \citenamefont {Mu{\~n}oz-Rojas}, \citenamefont {Nguyen}, \citenamefont {Moreau}, \citenamefont {Celle}, \citenamefont {Simonato}, \citenamefont {Br{\'e}chet},\ and\ \citenamefont {Bellet}}]{sannicolo_direct_2016}%
  \BibitemOpen
  \bibfield  {author} {\bibinfo {author} {\bibfnamefont {T.}~\bibnamefont {Sannicolo}}, \bibinfo {author} {\bibfnamefont {D.}~\bibnamefont {Mu{\~n}oz-Rojas}}, \bibinfo {author} {\bibfnamefont {N.~D.}\ \bibnamefont {Nguyen}}, \bibinfo {author} {\bibfnamefont {S.}~\bibnamefont {Moreau}}, \bibinfo {author} {\bibfnamefont {C.}~\bibnamefont {Celle}}, \bibinfo {author} {\bibfnamefont {J.-P.}\ \bibnamefont {Simonato}}, \bibinfo {author} {\bibfnamefont {Y.}~\bibnamefont {Br{\'e}chet}}, \ and\ \bibinfo {author} {\bibfnamefont {D.}~\bibnamefont {Bellet}},\ }\href {\doibase 10.1021/acs.nanolett.6b03270} {\bibfield  {journal} {\bibinfo  {journal} {Nano Letters}\ }\textbf {\bibinfo {volume} {16}},\ \bibinfo {pages} {7046} (\bibinfo {year} {2016})}\BibitemShut {NoStop}%
\bibitem [{\citenamefont {Diaz~Schneider}\ \emph {et~al.}(2022)\citenamefont {Diaz~Schneider}, \citenamefont {Angelom{\'e}}, \citenamefont {Granja}, \citenamefont {Quinteros}, \citenamefont {Levy},\ and\ \citenamefont {Mart{\'i}nez}}]{diaz_schneider_resistive_2022}%
  \BibitemOpen
  \bibfield  {author} {\bibinfo {author} {\bibfnamefont {J.~I.}\ \bibnamefont {Diaz~Schneider}}, \bibinfo {author} {\bibfnamefont {P.~C.}\ \bibnamefont {Angelom{\'e}}}, \bibinfo {author} {\bibfnamefont {L.~P.}\ \bibnamefont {Granja}}, \bibinfo {author} {\bibfnamefont {C.~P.}\ \bibnamefont {Quinteros}}, \bibinfo {author} {\bibfnamefont {P.~E.}\ \bibnamefont {Levy}}, \ and\ \bibinfo {author} {\bibfnamefont {E.~D.}\ \bibnamefont {Mart{\'i}nez}},\ }\href {\doibase 10.1002/aelm.202200631} {\bibfield  {journal} {\bibinfo  {journal} {Advanced Electronic Materials}\ }\textbf {\bibinfo {volume} {8}},\ \bibinfo {pages} {2200631} (\bibinfo {year} {2022})}\BibitemShut {NoStop}%
\bibitem [{\citenamefont {Di~Francesco}\ \emph {et~al.}(2021)\citenamefont {Di~Francesco}, \citenamefont {Sanca},\ and\ \citenamefont {Quinteros}}]{di_francesco_spatiotemporal_2021}%
  \BibitemOpen
  \bibfield  {author} {\bibinfo {author} {\bibfnamefont {F.}~\bibnamefont {Di~Francesco}}, \bibinfo {author} {\bibfnamefont {G.~A.}\ \bibnamefont {Sanca}}, \ and\ \bibinfo {author} {\bibfnamefont {C.~P.}\ \bibnamefont {Quinteros}},\ }\href {\doibase 10.1063/5.0067048} {\bibfield  {journal} {\bibinfo  {journal} {Applied Physics Letters}\ }\textbf {\bibinfo {volume} {119}},\ \bibinfo {pages} {193502} (\bibinfo {year} {2021})}\BibitemShut {NoStop}%
\bibitem [{\citenamefont {Chialvo}(2010)}]{chialvo_emergent_2010}%
  \BibitemOpen
  \bibfield  {author} {\bibinfo {author} {\bibfnamefont {D.~R.}\ \bibnamefont {Chialvo}},\ }\href {\doibase 10.1038/nphys1803} {\bibfield  {journal} {\bibinfo  {journal} {Nature Physics}\ }\textbf {\bibinfo {volume} {6}},\ \bibinfo {pages} {744} (\bibinfo {year} {2010})}\BibitemShut {NoStop}%
\bibitem [{\citenamefont {Diaz~Schneider}\ \emph {et~al.}(2024)\citenamefont {Diaz~Schneider}, \citenamefont {Quinteros}, \citenamefont {Levy},\ and\ \citenamefont {Mart{\'i}nez}}]{diaz_schneider_two-junction_nodate}%
  \BibitemOpen
  \bibfield  {author} {\bibinfo {author} {\bibfnamefont {J.~I.}\ \bibnamefont {Diaz~Schneider}}, \bibinfo {author} {\bibfnamefont {C.~P.}\ \bibnamefont {Quinteros}}, \bibinfo {author} {\bibfnamefont {P.}~\bibnamefont {Levy}}, \ and\ \bibinfo {author} {\bibfnamefont {E.~D.}\ \bibnamefont {Mart{\'i}nez}},\ }\href {\doibase 10.1002/adfm.202410766} {\bibfield  {journal} {\bibinfo  {journal} {Advanced Functional Materials}\ }\textbf {\bibinfo {volume} {n/a}},\ \bibinfo {pages} {2410766} (\bibinfo {year} {2024})}\BibitemShut {NoStop}%
\bibitem [{\citenamefont {Kendall}\ and\ \citenamefont {Kumar}(2020)}]{kendall_building_2020-1}%
  \BibitemOpen
  \bibfield  {author} {\bibinfo {author} {\bibfnamefont {J.~D.}\ \bibnamefont {Kendall}}\ and\ \bibinfo {author} {\bibfnamefont {S.}~\bibnamefont {Kumar}},\ }\href {\doibase 10.1063/1.5129306} {\bibfield  {journal} {\bibinfo  {journal} {Applied Physics Reviews}\ }\textbf {\bibinfo {volume} {7}},\ \bibinfo {pages} {011305} (\bibinfo {year} {2020})}\BibitemShut {NoStop}%
\bibitem [{\citenamefont {Jiu}\ \emph {et~al.}(2014)\citenamefont {Jiu}, \citenamefont {Araki}, \citenamefont {Wang}, \citenamefont {Nogi}, \citenamefont {Sugahara}, \citenamefont {Nagao}, \citenamefont {Koga}, \citenamefont {Suganuma}, \citenamefont {Nakazawa}, \citenamefont {Hara}, \citenamefont {Uchida},\ and\ \citenamefont {Shinozaki}}]{jiu_facile_2014}%
  \BibitemOpen
  \bibfield  {author} {\bibinfo {author} {\bibfnamefont {J.}~\bibnamefont {Jiu}}, \bibinfo {author} {\bibfnamefont {T.}~\bibnamefont {Araki}}, \bibinfo {author} {\bibfnamefont {J.}~\bibnamefont {Wang}}, \bibinfo {author} {\bibfnamefont {M.}~\bibnamefont {Nogi}}, \bibinfo {author} {\bibfnamefont {T.}~\bibnamefont {Sugahara}}, \bibinfo {author} {\bibfnamefont {S.}~\bibnamefont {Nagao}}, \bibinfo {author} {\bibfnamefont {H.}~\bibnamefont {Koga}}, \bibinfo {author} {\bibfnamefont {K.}~\bibnamefont {Suganuma}}, \bibinfo {author} {\bibfnamefont {E.}~\bibnamefont {Nakazawa}}, \bibinfo {author} {\bibfnamefont {M.}~\bibnamefont {Hara}}, \bibinfo {author} {\bibfnamefont {H.}~\bibnamefont {Uchida}}, \ and\ \bibinfo {author} {\bibfnamefont {K.}~\bibnamefont {Shinozaki}},\ }\href {\doibase 10.1039/C4TA00502C} {\bibfield  {journal} {\bibinfo  {journal} {Journal of Materials Chemistry A}\ }\textbf {\bibinfo {volume} {2}},\ \bibinfo {pages} {6326} (\bibinfo {year} {2014})}\BibitemShut {NoStop}%
\bibitem [{\citenamefont {Khanarian}\ \emph {et~al.}(2013)\citenamefont {Khanarian}, \citenamefont {Joo}, \citenamefont {Liu}, \citenamefont {Eastman}, \citenamefont {Werner}, \citenamefont {O'Connell},\ and\ \citenamefont {Trefonas}}]{khanarian_optical_2013}%
  \BibitemOpen
  \bibfield  {author} {\bibinfo {author} {\bibfnamefont {G.}~\bibnamefont {Khanarian}}, \bibinfo {author} {\bibfnamefont {J.}~\bibnamefont {Joo}}, \bibinfo {author} {\bibfnamefont {X.-Q.}\ \bibnamefont {Liu}}, \bibinfo {author} {\bibfnamefont {P.}~\bibnamefont {Eastman}}, \bibinfo {author} {\bibfnamefont {D.}~\bibnamefont {Werner}}, \bibinfo {author} {\bibfnamefont {K.}~\bibnamefont {O'Connell}}, \ and\ \bibinfo {author} {\bibfnamefont {P.}~\bibnamefont {Trefonas}},\ }\href {\doibase 10.1063/1.4812390} {\bibfield  {journal} {\bibinfo  {journal} {Journal of Applied Physics}\ }\textbf {\bibinfo {volume} {114}},\ \bibinfo {pages} {024302} (\bibinfo {year} {2013})}\BibitemShut {NoStop}%
\bibitem [{\citenamefont {Zhu}\ \emph {et~al.}(2019)\citenamefont {Zhu}, \citenamefont {Chen}, \citenamefont {Wan}, \citenamefont {Peng}, \citenamefont {Huang}, \citenamefont {Jiang}, \citenamefont {Li},\ and\ \citenamefont {Chu}}]{zhu_convertible_2019-1}%
  \BibitemOpen
  \bibfield  {author} {\bibinfo {author} {\bibfnamefont {Y.}~\bibnamefont {Zhu}}, \bibinfo {author} {\bibfnamefont {J.}~\bibnamefont {Chen}}, \bibinfo {author} {\bibfnamefont {T.}~\bibnamefont {Wan}}, \bibinfo {author} {\bibfnamefont {S.}~\bibnamefont {Peng}}, \bibinfo {author} {\bibfnamefont {S.}~\bibnamefont {Huang}}, \bibinfo {author} {\bibfnamefont {Y.}~\bibnamefont {Jiang}}, \bibinfo {author} {\bibfnamefont {S.}~\bibnamefont {Li}}, \ and\ \bibinfo {author} {\bibfnamefont {D.}~\bibnamefont {Chu}},\ }\href {\doibase 10.1021/acsaelm.9b00218} {\bibfield  {journal} {\bibinfo  {journal} {ACS Applied Electronic Materials}\ }\textbf {\bibinfo {volume} {1}},\ \bibinfo {pages} {1275} (\bibinfo {year} {2019})}\BibitemShut {NoStop}%
\bibitem [{\citenamefont {Venezian}(1994)}]{venezian_resistance_1994}%
  \BibitemOpen
  \bibfield  {author} {\bibinfo {author} {\bibfnamefont {G.}~\bibnamefont {Venezian}},\ }\href {\doibase 10.1119/1.17696} {\bibfield  {journal} {\bibinfo  {journal} {American Journal of Physics}\ }\textbf {\bibinfo {volume} {62}},\ \bibinfo {pages} {1000} (\bibinfo {year} {1994})}\BibitemShut {NoStop}%
\bibitem [{\citenamefont {Cserti}(2000)}]{cserti_application_2000}%
  \BibitemOpen
  \bibfield  {author} {\bibinfo {author} {\bibfnamefont {J.}~\bibnamefont {Cserti}},\ }\href {\doibase 10.1119/1.1285881} {\bibfield  {journal} {\bibinfo  {journal} {American Journal of Physics}\ }\textbf {\bibinfo {volume} {68}},\ \bibinfo {pages} {896} (\bibinfo {year} {2000})}\BibitemShut {NoStop}%
\bibitem [{\citenamefont {Bellew}\ \emph {et~al.}(2015)\citenamefont {Bellew}, \citenamefont {Manning}, \citenamefont {Gomes~da Rocha}, \citenamefont {Ferreira},\ and\ \citenamefont {Boland}}]{bellew_resistance_2015}%
  \BibitemOpen
  \bibfield  {author} {\bibinfo {author} {\bibfnamefont {A.~T.}\ \bibnamefont {Bellew}}, \bibinfo {author} {\bibfnamefont {H.~G.}\ \bibnamefont {Manning}}, \bibinfo {author} {\bibfnamefont {C.}~\bibnamefont {Gomes~da Rocha}}, \bibinfo {author} {\bibfnamefont {M.~S.}\ \bibnamefont {Ferreira}}, \ and\ \bibinfo {author} {\bibfnamefont {J.~J.}\ \bibnamefont {Boland}},\ }\href {\doibase 10.1021/acsnano.5b05469} {\bibfield  {journal} {\bibinfo  {journal} {ACS Nano}\ }\textbf {\bibinfo {volume} {9}},\ \bibinfo {pages} {11422} (\bibinfo {year} {2015})}\BibitemShut {NoStop}%
\bibitem [{\citenamefont {Sannicolo}\ \emph {et~al.}(2018)\citenamefont {Sannicolo}, \citenamefont {Charvin}, \citenamefont {Flandin}, \citenamefont {Kraus}, \citenamefont {Papanastasiou}, \citenamefont {Celle}, \citenamefont {Simonato}, \citenamefont {Mu{\~n}oz-Rojas}, \citenamefont {Jim{\'e}nez},\ and\ \citenamefont {Bellet}}]{sannicolo_electrical_2018}%
  \BibitemOpen
  \bibfield  {author} {\bibinfo {author} {\bibfnamefont {T.}~\bibnamefont {Sannicolo}}, \bibinfo {author} {\bibfnamefont {N.}~\bibnamefont {Charvin}}, \bibinfo {author} {\bibfnamefont {L.}~\bibnamefont {Flandin}}, \bibinfo {author} {\bibfnamefont {S.}~\bibnamefont {Kraus}}, \bibinfo {author} {\bibfnamefont {D.~T.}\ \bibnamefont {Papanastasiou}}, \bibinfo {author} {\bibfnamefont {C.}~\bibnamefont {Celle}}, \bibinfo {author} {\bibfnamefont {J.-P.}\ \bibnamefont {Simonato}}, \bibinfo {author} {\bibfnamefont {D.}~\bibnamefont {Mu{\~n}oz-Rojas}}, \bibinfo {author} {\bibfnamefont {C.}~\bibnamefont {Jim{\'e}nez}}, \ and\ \bibinfo {author} {\bibfnamefont {D.}~\bibnamefont {Bellet}},\ }\href {\doibase 10.1021/acsnano.8b01242} {\bibfield  {journal} {\bibinfo  {journal} {ACS Nano}\ }\textbf {\bibinfo {volume} {12}},\ \bibinfo {pages} {4648} (\bibinfo {year} {2018})}\BibitemShut {NoStop}%
\bibitem [{\citenamefont {Milano}\ \emph {et~al.}(2020)\citenamefont {Milano}, \citenamefont {Pedretti}, \citenamefont {Fretto}, \citenamefont {Boarino}, \citenamefont {Benfenati}, \citenamefont {Ielmini}, \citenamefont {Valov},\ and\ \citenamefont {Ricciardi}}]{milano_brain-inspired_2020}%
  \BibitemOpen
  \bibfield  {author} {\bibinfo {author} {\bibfnamefont {G.}~\bibnamefont {Milano}}, \bibinfo {author} {\bibfnamefont {G.}~\bibnamefont {Pedretti}}, \bibinfo {author} {\bibfnamefont {M.}~\bibnamefont {Fretto}}, \bibinfo {author} {\bibfnamefont {L.}~\bibnamefont {Boarino}}, \bibinfo {author} {\bibfnamefont {F.}~\bibnamefont {Benfenati}}, \bibinfo {author} {\bibfnamefont {D.}~\bibnamefont {Ielmini}}, \bibinfo {author} {\bibfnamefont {I.}~\bibnamefont {Valov}}, \ and\ \bibinfo {author} {\bibfnamefont {C.}~\bibnamefont {Ricciardi}},\ }\href {\doibase https://doi.org/10.1002/aisy.202000096} {\bibfield  {journal} {\bibinfo  {journal} {Advanced Intelligent Systems}\ }\textbf {\bibinfo {volume} {2}},\ \bibinfo {pages} {2000096} (\bibinfo {year} {2020})}\BibitemShut {NoStop}%
\bibitem [{\citenamefont {Milano}\ \emph {et~al.}(2024)\citenamefont {Milano}, \citenamefont {Raffone}, \citenamefont {Bejtka}, \citenamefont {Carlo}, \citenamefont {Fretto}, \citenamefont {Pirri}, \citenamefont {Cicero}, \citenamefont {Ricciardi},\ and\ \citenamefont {Valov}}]{milano_electrochemical_2024-1}%
  \BibitemOpen
  \bibfield  {author} {\bibinfo {author} {\bibfnamefont {G.}~\bibnamefont {Milano}}, \bibinfo {author} {\bibfnamefont {F.}~\bibnamefont {Raffone}}, \bibinfo {author} {\bibfnamefont {K.}~\bibnamefont {Bejtka}}, \bibinfo {author} {\bibfnamefont {I.~D.}\ \bibnamefont {Carlo}}, \bibinfo {author} {\bibfnamefont {M.}~\bibnamefont {Fretto}}, \bibinfo {author} {\bibfnamefont {F.~C.}\ \bibnamefont {Pirri}}, \bibinfo {author} {\bibfnamefont {G.}~\bibnamefont {Cicero}}, \bibinfo {author} {\bibfnamefont {C.}~\bibnamefont {Ricciardi}}, \ and\ \bibinfo {author} {\bibfnamefont {I.}~\bibnamefont {Valov}},\ }\href {\doibase 10.1039/D3NH00476G} {\bibfield  {journal} {\bibinfo  {journal} {Nanoscale Horizons}\ }\textbf {\bibinfo {volume} {9}},\ \bibinfo {pages} {416} (\bibinfo {year} {2024})}\BibitemShut {NoStop}%
\bibitem [{\citenamefont {Koo}\ \emph {et~al.}(2021)\citenamefont {Koo}, \citenamefont {Park}, \citenamefont {Koo},\ and\ \citenamefont {Kim}}]{koo_local_2021}%
  \BibitemOpen
  \bibfield  {author} {\bibinfo {author} {\bibfnamefont {S.}~\bibnamefont {Koo}}, \bibinfo {author} {\bibfnamefont {J.}~\bibnamefont {Park}}, \bibinfo {author} {\bibfnamefont {S.}~\bibnamefont {Koo}}, \ and\ \bibinfo {author} {\bibfnamefont {K.}~\bibnamefont {Kim}},\ }\href {\doibase 10.1021/acs.jpcc.0c10774} {\bibfield  {journal} {\bibinfo  {journal} {The Journal of Physical Chemistry C}\ }\textbf {\bibinfo {volume} {125}},\ \bibinfo {pages} {6306} (\bibinfo {year} {2021})}\BibitemShut {NoStop}%
\bibitem [{Note1()}]{Note1}%
  \BibitemOpen
  \bibinfo {note} {This is related to the areal density of NWs and the lateral dimensions of the electrodes.}\BibitemShut {Stop}%
\bibitem [{Note2()}]{Note2}%
  \BibitemOpen
  \bibinfo {note} {Irreversibility may arise on our experimental conditions (electrodes' separation, SANs density, and/or applied electrical protocol). Other authors have shown a more controlled disruption for individual NWs that can be rewired back upon proper electrical stimulation \cite {milano_electrochemical_2024-1}.}\BibitemShut {Stop}%
\bibitem [{\citenamefont {Kholid}\ \emph {et~al.}(2015)\citenamefont {Kholid}, \citenamefont {Huang}, \citenamefont {Zhang},\ and\ \citenamefont {Fan}}]{kholid_multiple_2015}%
  \BibitemOpen
  \bibfield  {author} {\bibinfo {author} {\bibfnamefont {F.~N.}\ \bibnamefont {Kholid}}, \bibinfo {author} {\bibfnamefont {H.}~\bibnamefont {Huang}}, \bibinfo {author} {\bibfnamefont {Y.}~\bibnamefont {Zhang}}, \ and\ \bibinfo {author} {\bibfnamefont {H.~J.}\ \bibnamefont {Fan}},\ }\href {\doibase 10.1088/0957-4484/27/2/025703} {\bibfield  {journal} {\bibinfo  {journal} {Nanotechnology}\ }\textbf {\bibinfo {volume} {27}},\ \bibinfo {pages} {025703} (\bibinfo {year} {2015})}\BibitemShut {NoStop}%
\bibitem [{\citenamefont {Yang}\ \emph {et~al.}(2014)\citenamefont {Yang}, \citenamefont {Gao}, \citenamefont {Li}, \citenamefont {Pan}, \citenamefont {Tappertzhofen}, \citenamefont {Choi}, \citenamefont {Waser}, \citenamefont {Valov},\ and\ \citenamefont {Lu}}]{yang_electrochemical_2014}%
  \BibitemOpen
  \bibfield  {author} {\bibinfo {author} {\bibfnamefont {Y.}~\bibnamefont {Yang}}, \bibinfo {author} {\bibfnamefont {P.}~\bibnamefont {Gao}}, \bibinfo {author} {\bibfnamefont {L.}~\bibnamefont {Li}}, \bibinfo {author} {\bibfnamefont {X.}~\bibnamefont {Pan}}, \bibinfo {author} {\bibfnamefont {S.}~\bibnamefont {Tappertzhofen}}, \bibinfo {author} {\bibfnamefont {S.}~\bibnamefont {Choi}}, \bibinfo {author} {\bibfnamefont {R.}~\bibnamefont {Waser}}, \bibinfo {author} {\bibfnamefont {I.}~\bibnamefont {Valov}}, \ and\ \bibinfo {author} {\bibfnamefont {W.~D.}\ \bibnamefont {Lu}},\ }\href {\doibase 10.1038/ncomms5232} {\bibfield  {journal} {\bibinfo  {journal} {Nature Communications}\ }\textbf {\bibinfo {volume} {5}},\ \bibinfo {pages} {4232} (\bibinfo {year} {2014})}\BibitemShut {NoStop}%
\bibitem [{\citenamefont {Wang}\ \emph {et~al.}(2019)\citenamefont {Wang}, \citenamefont {Wang}, \citenamefont {Ambrosi}, \citenamefont {Bricalli}, \citenamefont {Laudato}, \citenamefont {Sun}, \citenamefont {Chen},\ and\ \citenamefont {Ielmini}}]{wang_surface_2019-1}%
  \BibitemOpen
  \bibfield  {author} {\bibinfo {author} {\bibfnamefont {W.}~\bibnamefont {Wang}}, \bibinfo {author} {\bibfnamefont {M.}~\bibnamefont {Wang}}, \bibinfo {author} {\bibfnamefont {E.}~\bibnamefont {Ambrosi}}, \bibinfo {author} {\bibfnamefont {A.}~\bibnamefont {Bricalli}}, \bibinfo {author} {\bibfnamefont {M.}~\bibnamefont {Laudato}}, \bibinfo {author} {\bibfnamefont {Z.}~\bibnamefont {Sun}}, \bibinfo {author} {\bibfnamefont {X.}~\bibnamefont {Chen}}, \ and\ \bibinfo {author} {\bibfnamefont {D.}~\bibnamefont {Ielmini}},\ }\href {\doibase 10.1038/s41467-018-07979-0} {\bibfield  {journal} {\bibinfo  {journal} {Nat Commun}\ }\textbf {\bibinfo {volume} {10}},\ \bibinfo {pages} {81} (\bibinfo {year} {2019})}\BibitemShut {NoStop}%
\bibitem [{\citenamefont {Daniels}\ \emph {et~al.}(2022)\citenamefont {Daniels}, \citenamefont {Mallinson}, \citenamefont {Heywood}, \citenamefont {Bones}, \citenamefont {Arnold},\ and\ \citenamefont {Brown}}]{daniels_reservoir_2022}%
  \BibitemOpen
  \bibfield  {author} {\bibinfo {author} {\bibfnamefont {R.~K.}\ \bibnamefont {Daniels}}, \bibinfo {author} {\bibfnamefont {J.~B.}\ \bibnamefont {Mallinson}}, \bibinfo {author} {\bibfnamefont {Z.~E.}\ \bibnamefont {Heywood}}, \bibinfo {author} {\bibfnamefont {P.~J.}\ \bibnamefont {Bones}}, \bibinfo {author} {\bibfnamefont {M.~D.}\ \bibnamefont {Arnold}}, \ and\ \bibinfo {author} {\bibfnamefont {S.~A.}\ \bibnamefont {Brown}},\ }\href {\doibase 10.1016/j.neunet.2022.07.001} {\bibfield  {journal} {\bibinfo  {journal} {Neural Networks}\ }\textbf {\bibinfo {volume} {154}},\ \bibinfo {pages} {122} (\bibinfo {year} {2022})}\BibitemShut {NoStop}%
\bibitem [{\citenamefont {Dayan}\ and\ \citenamefont {Abbott}(2005)}]{dayan_theoretical_2005}%
  \BibitemOpen
  \bibfield  {author} {\bibinfo {author} {\bibfnamefont {P.}~\bibnamefont {Dayan}}\ and\ \bibinfo {author} {\bibfnamefont {L.~F.}\ \bibnamefont {Abbott}},\ }\href@noop {}\ (\bibinfo  {publisher} {MIT Press},\ \bibinfo {year} {2005})\BibitemShut {NoStop}%
\bibitem [{\citenamefont {Zhu}\ \emph {et~al.}(2021)\citenamefont {Zhu}, \citenamefont {Hochstetter}, \citenamefont {Loeffler}, \citenamefont {Diaz-Alvarez}, \citenamefont {Nakayama}, \citenamefont {Lizier},\ and\ \citenamefont {Kuncic}}]{zhu_information_2021}%
  \BibitemOpen
  \bibfield  {author} {\bibinfo {author} {\bibfnamefont {R.}~\bibnamefont {Zhu}}, \bibinfo {author} {\bibfnamefont {J.}~\bibnamefont {Hochstetter}}, \bibinfo {author} {\bibfnamefont {A.}~\bibnamefont {Loeffler}}, \bibinfo {author} {\bibfnamefont {A.}~\bibnamefont {Diaz-Alvarez}}, \bibinfo {author} {\bibfnamefont {T.}~\bibnamefont {Nakayama}}, \bibinfo {author} {\bibfnamefont {J.~T.}\ \bibnamefont {Lizier}}, \ and\ \bibinfo {author} {\bibfnamefont {Z.}~\bibnamefont {Kuncic}},\ }\href {\doibase 10.1038/s41598-021-92170-7} {\bibfield  {journal} {\bibinfo  {journal} {Scientific Reports}\ }\textbf {\bibinfo {volume} {11}},\ \bibinfo {pages} {13047} (\bibinfo {year} {2021})}\BibitemShut {NoStop}%
\end{thebibliography}%

\end{document}